\documentclass[a4paper,journal]{IEEEtran}

\IEEEoverridecommandlockouts

\newcommand\hmmax{0}
\newcommand\bmmax{0}

\usepackage{graphicx}
\usepackage{tikz}
\usepackage{pgfplots}
\usepackage{xcolor}
\usepackage{color, colortbl}
\usepackage{amsmath}
\usepackage{bbm}
\usepackage{amsfonts, amssymb}
\usepackage{mathtools}
\usepackage[pdftex,pdfauthor={Kristian Skafte Jensen},pdftitle={Classical Communication Protocol based on Joint Classical-Quantum Coding}]{hyperref} 		
\usepackage{cite}
\usepackage[nolist]{acronym}
\usepackage{amsmath}

\usepackage{placeins}
\usepackage{booktabs} 		
\usepackage{diagbox}		
\usepackage{upgreek}
\usepackage{bbm}
\usepackage{svg}
\usepackage{Files/bm}
\usepackage{Files/winsnotation}
\usepackage{Files/Symbols_OPL}
\usepackage{Files/LVcolours} 

\usepackage{listings}
\usepackage{Files/lstlang0}

\definecolor{codegreen}{rgb}{0,0.6,0}
\definecolor{codegray}{rgb}{0.5,0.5,0.5}
\definecolor{codepurple}{rgb}{0.58,0,0.82}
\definecolor{backcolour}{rgb}{0.95,0.95,0.92}

\lstdefinestyle{mystyle}{
    backgroundcolor=\color{backcolour},   
    commentstyle=\color{codegreen},
    keywordstyle=\color{magenta},
    numberstyle=\tiny\color{codegray},
    stringstyle=\color{codepurple},
    basicstyle=\ttfamily\footnotesize,
    breakatwhitespace=false,         
    breaklines=true,                 
    captionpos=b,                    
    keepspaces=true,                 
    numbers=left,                    
    numbersep=5pt,                  
    showspaces=false,                
    showstringspaces=false,
    showtabs=false,                  
    tabsize=2
}

\usepackage[ruled,vlined,noresetcount]{algorithm2e}
\usepackage{setspace}	 	
\usepackage{comment}
\usepackage{physics}
\usepackage{multirow}
\usepackage{ragged2e}

\usepackage{soul}

\usepackage{array}
\newcolumntype{C}[1]{>{\centering\arraybackslash}p{#1}}

\makeatletter
\def\endthebibliography{%
  \def\@noitemerr{\@latex@warning{Empty `thebibliography' environment}}%
  \endlist
}
\makeatother
\usepackage{amsthm}
\theoremstyle{definition}

\pgfplotsset{compat=1.17}

\newcommand{\C}{\mathcal{C}}

\newcommand{\Cout}{\mathcal{C}_{\mathrm{out}}}
\newcommand{\Cin}{\mathcal{C}_{\mathrm{in}}}
\newcommand{\nout}{{n_{\mathrm{out}}}}
\newcommand{\nin}{{n_{\mathrm{in}}}}
\newcommand{\kout}{{k_{\mathrm{out}}}}
\newcommand{\kin}{{k_{\mathrm{in}}}}
\newcommand{\dout}{{d_{\mathrm{out}}}}
\newcommand{\din}{{d_{\mathrm{in}}}}

\begin{document}


\title{
Classical Communication Protocol based on \\Joint Classical-Quantum  Coding
}

\author{%
    \IEEEauthorblockN{ Kristian~Skafte~Jensen, René~Bødker~Christensen, \v Cedomir~Stefanovi\' c, Petar~Popovski}
    \thanks{Kristian~Skafte~Jensen, \v Cedomir~Stefanovi\' c and Petar~Popovski are with Department of Electronic Systems, Aalborg University, Denmark. Email: \{ksjen, cs, petarp\}@es.aau.dk}
    \thanks{René~Bødker~Christensen is with Department of Mathematical Sciences, Aalborg University, Denmark. Email: rene@math.aau.dk}
    \thanks{This work was supported in part by the Danish National Research Foundation through the CLASSIQUE Center, under Grant 187.}
}

\maketitle 

\begin{acronym}
\small
\acro{AWGN}{additive white Gaussian noise}
\acro{BCH}{Bose–Chaudhuri–Hocquenghem}
\acro{CDF}{cumulative distribution function}
\acro{CRC}{cyclic redundancy code}
\acro{CSS}{Calderbank-Shor-Steane}
\acro{EPR}{Einstein–Podolsky–Rosen}
\acro{i.i.d.}{independent and identically distributed}
\acro{LDPC}{low-density parity-check}
\acro{ML}{maximum likelihood}
\acro{MWPM}{minimum weight perfect matching}
\acro{QECC}{quantum error correcting code}
\acro{QEC}{quantum error correction}
\acro{QKD}{quantum key distribution}
\acro{PDF}{probability density function}
\acro{PMF}{probability mass function}
\acro{MPS}{matrix product state}
\acro{WB}{well-behaving}
\acro{WEP}{weight enumerator polynomial}
\acro{WE}{weight enumerator}
\end{acronym}
\setcounter{page}{1}

\begin{abstract}
    We introduce a robust quantum communication protocol that  integrates classical error-control coding, entanglement distribution, and superdense coding. Classical error-correcting codes are used to mitigate dark counts and photon losses by  determining the positions of qubit transmissions and protecting the data embedded through superdense coding. We derive conditions on the employed codes that guarantee successful error correction under a bounded error-frequency model. Moreover, upper bounds are derived on the performance of conventional superdense coding protected by classical error correction. It is shown that, 
    under the same constraints on error frequency, suitable code configurations of the proposed protocol can exceed those upper bounds both in terms of data rate and energy efficiency.  Finally, we develop a physical error model based on fiber attenuation, detector efficiency, dark counts, and time-slot duration, and use it to evaluate the effective performance of different configurations of error-correcting codes. The proposed approach is primarily suited for short-distance quantum links, as in Quantum Local Area Network (QLAN) where it can provide high communication throughput while integrating entanglement distribution directly into the communication process.
\end{abstract}

\begin{keywords}
    Quantum communication, classical error correction, entanglement distribution, superdense coding
\end{keywords}

\section{Introduction}

Superdense coding \cite{BenWie92:SuperDenseCoding} is a well-known technique to send classical bits via a quantum channel. More precisely, two classical bits are encoded in the transmission of a single qubit at the cost of one pre-shared EPR pair. While superdense coding has a conceptual value that reveals the features of quantum entanglement, sending multiple bits with a single symbol is routinely done in classical communication. For instance, Quadrature Phase Shift Keying (QPSK) \cite{QPSK} is a classical modulation technique that sends two bits by encoding them into one of the four phases of the carrier. QPSK requires phase synchronization, typically carried out at the start of the packet and then reused for all symbols. Phase synchronization for QPSK is functionally identical to the preshared entanglement in superdense coding. A crucial difference between the two, however, is the fact that the entanglement is destroyed upon a single 2-bit transmission through superdense coding. This would correspond to a scenario in which the phase synchronization in QPSK is lost upon the decoding of each symbol, rendering the QPSK-based systems unusable in practice.

As shown in~\cite{JensenEtAl25}, the challenge of generating EPR pairs without losing too much communication throughput may be solved by encoding information both via superdense coding, but also via the positions of the qubit transmissions.
In brief, the presence of a qubit encodes a classical `1', and the absence encodes a classical `0'.
In this way, EPR generation is tied directly to the communication protocol instead of being assumed to be simply an available resource.
However, the main advantage introduced in this scheme arrives from the introduction of an intentional Pauli error on the last qubit of each pair, which encodes two additional classical bits. 
This instance of superdense coding can be seen as piggybacking of classical data over a quantum carrier \cite{ChiConWin:20}.
The drawback of the scheme is that it can operate only on the error-free quantum channel, which is a requirement seldom met in practice.

\begin{figure*}[t]
    \centering
    \includegraphics[width=0.8\linewidth]{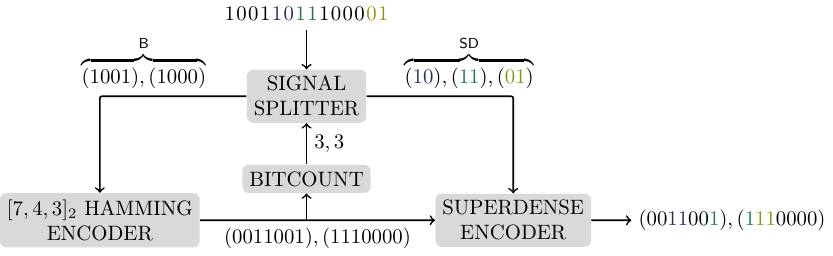}
    \caption{Illustrative example of the idea behind the proposed encoding scheme. The string of information bits is fed into the signal splitter which encodes chunks of length $k$ given some $[n,k,d]$ code. 
    The '1's from the codeword are transmitted using qubits, the pairs of which are entangled and superdense encoding, representing additional bits from the input stream.
    }
    \label{fig:blockdiagram}
\end{figure*}

In this work, we develop a protocol that leverages the idea of combining classical and quantum coding and is able to combat the impairments of the optical channel, namely \emph{dark counts} and \emph{photon losses}. 
The developed protocol builds up on the ideas presented in~\cite{JensenEtAl25}, introducing a non-trivial extension that employs classical error correction methods to counteract potential errors in data encoded via superdense coding. A visualization of the main idea behind the proposed encoding approach can be seen in Fig.~\ref{fig:blockdiagram}.

We show that, under the same constraints on allowed error frequency, our protocol yields improved performance, both in terms of data throughput and energy efficiency, than merely applying classical error correction on regular superdense coding. Specifically, we derive upper bounds on the performance of error-corrected superdense coding and show that a specific choice of error-correcting codes used with our proposed protocol exceeds the performance of superdense coding no matter the choice of codes applied to it for the same constraints on error frequency. The proposed method is particularly suitable for short-range optical interconnects in distributed computing environments~\cite{Cacciapuoti2026} and quantum data centers~\cite{Shapourian2025QDC}.

The rest of the paper is organized as follows.
Section~\ref{sec:rw} provides a brief introduction to the established work that the current proposal builds on.
In Section \ref{sec:model}, we characterize the types of errors that occur in the channel and determine the sources of these errors. We then proceed to present a model within which we safeguard the communication aspect of the protocol utilizing classical error-correcting codes. To this end, we derive the performance metrics for each error model. In Section \ref{sec:handl-prot-errors}, the treatment of the channel is split into two separate scenarios to get a better grasp of the distinctions between the error types and how to effectively combat them. Clarifying examples of the proposed protocols are given. In Section \ref{sec:modelling-error-prob}, we estimate error probabilities based on physical parameters obtained in experimental work in the literature, which leads to model simulations in Section \ref{sec:results}. Finally, in Section~\ref{sec:discussion} we discuss the results and their implications for our proposal.

\section{Related work}
\label{sec:rw}

Superdense coding is a method for encoding classical data into qubits in order to essentially double the data throughput \cite{BenWie92:SuperDenseCoding}. Suppose Alice wishes to transmit a message to Bob, but she is limited to sending a single unit of data at a time. Classically, this would restrict the rate to 1 bit per time unit. Now, assume that Alice and Bob are capable of generating, transmitting, and measuring qubits. Further assume that Alice and Bob share a supply of Einstein-Podolsky-Rosen (EPR) pairs \cite{Einstein35}. Before transmitting a qubit, Alice encodes classical information onto the EPR pair by applying an operator from the basis $\{I,X,Y,Z\}$ on her qubit. With four unique outcomes, this allows for 2 bits of classical data, i.e., 1 classical bit per generated qubit of the pair. After receiving the qubit, Bob performs projective measurements on the EPR pair composed in part by the qubit received from Alice to obtain the classical data.

A technique proposed in \cite{ChiConWin:20} allows classical information to be carried by quantum packets, constituting a form of piggybacking. The idea relies on quantum error correction, and specifically on the intentional employment of errors on codewords. In short, Alice utilizes an $[\![n,k]\!]$ quantum error-correcting code to encode $k$ data qubits into a codeword of $n$ qubits. She then applies to the codeword operators corresponding to correctable errors. After receiving the modified codeword, Bob measures the syndrome and can thus deduce which intentional errors were applied and in which order. The presence of specific, correctable errors acts as encoding of classical information.

The idea of piggybacking classical information over quantum channels was developed further in \cite{JensenEtAl25}.
The scheme assumes a time-slotted channel, where the transmission of a qubit in a slot encodes a classical `1', and the absence of a qubit transmission encodes a classical `0'. Moreover, pairs of consecutively transmitted qubits constitute entangled, EPR pairs, where two additional classical bits are encoded by inducing an intentional Pauli error on the last qubit.
In this way, we are able to transmit three information bits in a single slot: one bit due to the protocol (the `1' which triggers the qubit transmission) and two bits due to superdense coding over an EPR pair.
The protocol is effectively structured into rounds, where each round involves a user transmitting a sequence $[S_1, 1, S_2, 1]$, where $S_1$ and $S_2$ are sequences of zeroes of arbitrary length.

The performance of the scheme in~\cite{JensenEtAl25} depends on the coherence time $c$ of the generated EPR pairs. Assuming that the source is memoryless and generates classical 0's and 1's with equal probability, it was shown in~\cite{JensenEtAl25} that the communication rate (measured in classical bits per time slot) is $4/3$ in the extreme case where EPR pairs remain coherent for only $c=2$ time slots (i.e., the length of the sequence $S_2$ is 0). In other words, in this case both qubits have to be transmitted immediately one after the other; otherwise they will decohere.
As $c$ increases, it was shown that the rate converges to $3/2$.

\section{System model and performance metrics}
\label{sec:model}

Consider a simple setup consisting of two users, $U_1$ and $U_2$, in which $U_1$ wants to send classical information from an infinitely long bit stream to $U_2$ across separate time slots using the transmission of qubits, with no more than one transmission in each time slot.
At any given time slot, $U_1$ has the following options:
\begin{itemize}
    \item Do nothing,
    \item Generate an EPR pair and transmit one qubit,
    \item Apply superdense coding on an already shared EPR pair, by performing a Pauli operation on the second qubit of the pair.
\end{itemize}
As already noted, the absence of a transmission (doing nothing) encodes 0, transmission of  qubit encodes 1, and superdense coding encodes two bits from the bit stream.  
In order to properly design a protocol that leverages the above operations and that is reliable against errors, it is necessary to first consider the types of physical errors that may happen and how to appropriately model these during the analysis.
We present these considerations in the following sections, along with the metrics that we will use to evaluate the performance of our proposed solutions.

\subsection{Error sources and error types}
\label{sec:error-types-quantum}

We consider two sources of physical errors in the quantum channel:
\begin{itemize}
  \item \textit{Dark counts:} The photon detector may signal the arrival of a photon despite no actual photon being present. Additionally, rogue photons from the environment might enter the optical fibre. In both cases, the transmission of a `0' is read as a `1' by the receiver.
  \item \textit{Photon losses:} Due to attenuation, a transmitted photon is lost in the optical fibre, and detectors might fail to detect an incoming photon. This leads the receiver to assume that a `0' was transmitted rather than a `1'.
\end{itemize}
These physical errors will affect the data bits, either in the sequence of transmitted qubits or in the bits encoded using superdense coding. Both dark counts and photon losses may be seen as bit flips on the bits encoded via qubit transmission.
When it comes to the bits encoded via superdense coding, a photon loss is the critical error type since one half of an EPR pair is lost, and the usual measurements of superdense coding cannot be performed. One may see this as an erasure of the corresponding superdense-encoded data bits.  
We will refer to these bit flip errors and erasures as \emph{data bit errors}.

Since qubits are prepared in pairs, but transmitted individually, both dark counts and photon losses will cause a desynchronization between sender and receiver. That is, if no error-correction is performed, the receiver may end up measuring qubits from different EPR pairs (in case of a photon loss) or even unentangled photons (in case of a dark count) when trying to extract the information encoded via superdense coding.
We refer to this type of error as a \emph{protocol error} since they affect the structure of the protocol.

In practice, another source of errors can be the imperfect generation of EPR pairs. This means that the generated entanglement will have reduced fidelity, which may cause data bit errors when using them for superdense coding.
In this paper, we assume perfect EPR pair generation and focus on solving data bit and protocol errors.

\subsection{Error model}
\label{sec:error-model}

Throughout the work, we assume that no more than $1$ error (either dark count or photon loss) occurs for every $N$ consecutive time slots.
$N$ will depend on the technology used, and it dictates the types of error-correction scheme that can be applied successfully. Clearly, higher values of $N$ impose fewer restrictions on the scheme used. Alternatively, one may focus on a specific correction scheme and derive the minimal $N$ from this. In this case, $N$ represents minimal requirements on the technology used to implement the quantum channel.

Furthermore, we adopt the same decoherence model as in~\cite{JensenEtAl25}, such that EPR pairs are assumed to be perfectly coherent for $c$ time slots for some integer $c\geq 2$.

\subsection{Performance metrics}

To gauge the performance of the proposed protocol, we adopt the following metrics.
The data rate $R$ is defined as the number of transmitted information bits per timeslot.
More precisely, if we split the transmissions into rounds of $n$ timeslots, the data rate is given by
\begin{align}
  R= \lim_{\ell\rightarrow\infty} \frac{\sum_{i=1}^\ell \mathsf{B}_i}{n\ell},
\end{align}
where $\mathsf{B}_i$ denotes the number of transmitted information bits in the $i$'th round of $n$ timeslots. As a secondary metric, we use the energy efficiency $E$, which is defined as the number of transmitted information bits per physical qubit transmission:
\begin{equation}\label{eq:defEfficiency}
    E = \lim_{\ell\to\infty}\frac{\sum_{i=1}^\ell\mathsf{B}_i}{\sum_{i=1}^\ell\mathsf{Tx}_i},
\end{equation}
where $\mathsf{B}_i$ is as before and $\mathsf{Tx}_i$ denotes the number of physical transmissions (i.e., of transmitted photons) within the $i$'th round.

\section{Proposed protocol}
\label{sec:handl-prot-errors}
We propose a solution that employs classical error-correcting codes to protect the data transmission against physical errors (i.e., against bit flips and erasures) in the channel.

The bit stream is split into bits that will be used for classical encoding and superdense encoding.
Specifically, the sender $U_1$ utilizes an $[n,k,d]$ classical error-correcting code and encodes chunks of $k$ bits into length $n$ codewords. 
The sender records the number of `1's contained in the produced codeword (i.e., its Hamming weight), thus keeping track of the number of qubits that will be sent.
To utilize superdense coding, each pair of these qubits
will constitute an EPR pair prepared and superdense encoded by $U_1$.

In general, the number of the qubits to be transmitted in a codeword may be odd.
In this case, the final `1' of the codeword corresponds to one half of an EPR pair that will then be ``coupled'' with the first qubit of the next non-zero codeword.\footnote{Also, recall that the superdense encoding is performed solely on the second qubit in the EPR pair.}
For example, the code used in Fig.~\ref{fig:blockdiagram} is the $[7,4,3]$ Hamming code, so the information stream sends chunks of four bits to the encoder. In the depicted example, $1001$ is the first chunk which provides the codeword $0011001$. This codeword has Hamming weight 3, and the first two qubits constitute an EPR pair that superdense encodes the next two bits from the input stream (10 denoted in magenta in the figure, which are ``modulated'' into pair 11 in magenta at the superdense encoder's output). The sender then prepares the next EPR pair, with the third qubit of the codeword $0011001$ constituting the first qubit of the pair, and the next two bits (denoted by 11 in green) will be reserved for superdense encoding.

The next four bits from the input stream, which are 1000, are then encoded by the Hamming code, producing the codeword 1110000 that contains three qubits. The first qubit of this codeword is the second qubit of the pair that will superdense encode the reserved 11 in green. The next two qubits make an EPR pair that will superdense encode 01 denoted in yellow.
The protocol operation continues along these lines.

In the rest of the paper, we will abstract away the signal splitter in Fig.~\ref{fig:blockdiagram} and simply assume that the information stream of $U_1$ is split into two: a stream $\mathsf{B}$ allocated to encoding via a classical error-correcting code and a stream $\mathsf{SD}$ allocated to encoding via superdense coding.

\subsection{Correcting a single dark count}\label{sec:corr-single-dark}
We first consider the simplest case where the type of error is restricted to dark counts.

\begin{figure*}[!t]
    \centering
    \includegraphics[width=0.95\linewidth]{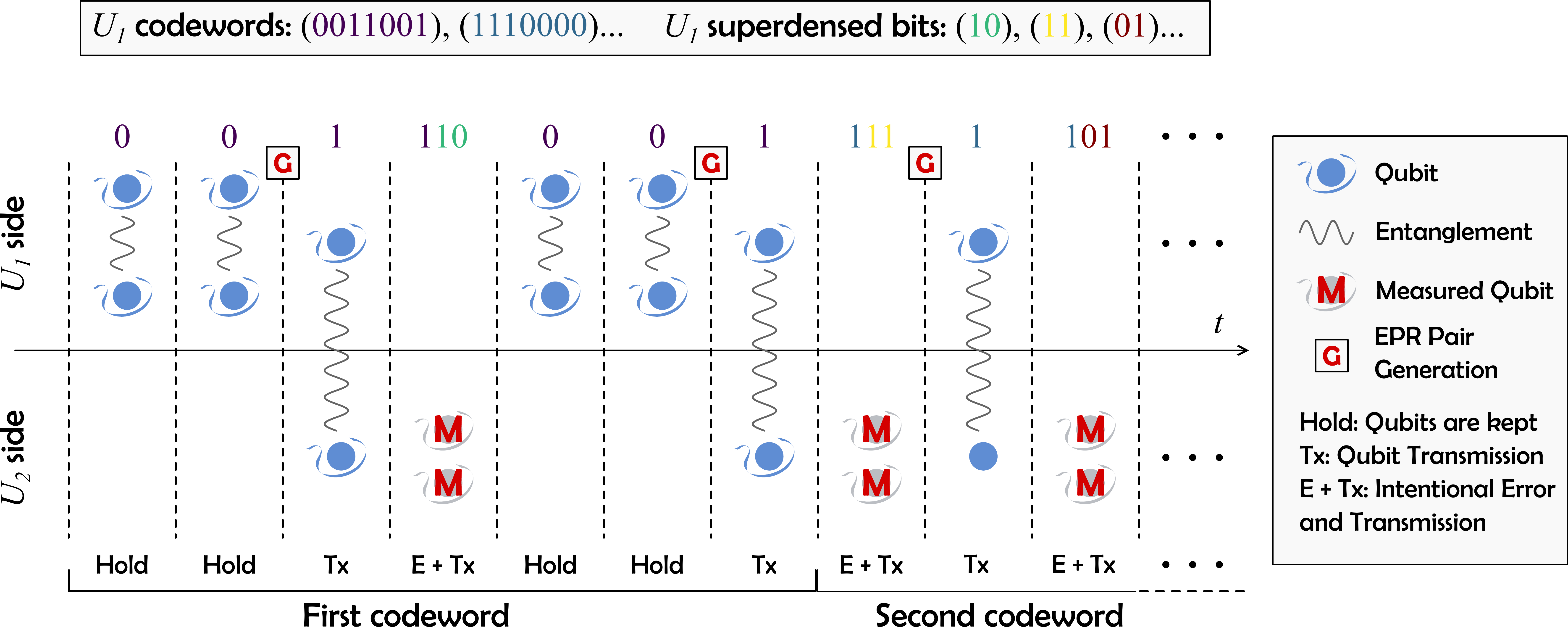}
    \caption{Proposed protocol to manage bit flip errors caused by dark counts and photon losses using classical error-correcting codes. The transmitting user $U_1$ communicates bits to the receiving user $U_2$, by sending the qubits of an EPR pair through a quantum channel, in a slotted setting. These bits are determined by codewords generated as in Fig.~\ref{fig:blockdiagram}. The quantum transmission occurs each time a `1' appears in the codeword. When the second qubit of the pair has to be transmitted, two extra information bits are superdense encoded in the EPR pair to enhance the data throughput.}
    \label{fig:2WayProtocol}
\end{figure*}

As described previously, we introduce a classical error-correcting code with parameters $[n,k,d]$.
The codewords of the code will dictate the qubit transmissions, and if the code parameters are chosen appropriately, this will allow detection and correction of the dark counts.
More precisely, the transmitting user will encode $k$ bits from $\mathsf{B}$ to a codeword of $n$ bits. These $n$ bits determine the transmission of qubits.
As in the error-free case, `0' translates to an empty time slot and `1' translates to transmission of a qubit. An example of this can be seen in Fig.~\ref{fig:2WayProtocol}.

By assumption, any transmission of a codeword in this way can be affected by at most $e=\lceil n/N\rceil$ dark counts. The dark counts correspond exactly to classical errors, meaning that they are correctable as long as $e<d/2$.

The proposed approach requires extra care when codewords have odd weight since EPR pairs are split across multiple transmission rounds in this case, necessitating support for potentially long coherence times.
Specifically, if the technology allows it, an option is to store the final qubit in memory until the second half of the EPR pair arrives in the subsequent transmission round(s). 

A straightforward solution to this challenge is to discard the second qubit of the final EPR pair if it cannot be transmitted in the same round.
Another solution is to use codes where all codewords have even weight.\footnote{Note that such codes are easily constructed by adding a parity bit to each codeword.} A special case of this are the constant composition codes.
Both of these ensure that all EPR pairs are prepared and consumed within a single transmission round, and hence need only stay coherent for one such round. This comes at the cost of sacrificing some of the transmission rate; the former approach by reducing the contribution to the overall transmission rate by the superdense coding, the latter by reducing throughput via qubit transmission. 

To analyse the rates of these two solutions, assume that an $[n,k,d]$ code $\mathcal{C}$ is used with weight distribution $A(\mathcal{C})=(A_0,A_1,\ldots,A_n)$. In the first case, i.e., when the second qubit of the pair is discarded as it cannot be transmitted in the same codeword, the transmission rate $R$ is then given by
\begin{align}\label{eq:rateDiscardHalfpairs}
  R &= \frac{1}{n}\left(k + \frac{1}{2^k}\sum_{i=0}^n 2\left\lfloor\frac{i}{2}\right\rfloor A_i\right)\nonumber\\
    &= \frac{k}{n} + \frac{1}{n2^{k-1}}\sum_{i=0}^n\left\lfloor\frac{i}{2}\right\rfloor A_i
\end{align}
where the first term on the right corresponds to the ``classical'' code rate and the second term corresponds to the increase in the code rate through the superdense coding.
Similarly, the efficiency is given by
\begin{equation}
    E = \frac{k+1/2^k\sum_{i=0}^n 2\lfloor i/2\rfloor A_i}{1/2^k\sum_{i=0}^niA_i}.
\end{equation}
One can show that for all binary linear codes of odd length, we have
\begin{equation*}
    \frac{1}{2^k}\sum_{i=0}^niA_i = \frac{n}{2},
\end{equation*}
thus for such codes it holds that $E=2R$.

If $\mathcal{C}$ is even-weight, then~\eqref{eq:rateDiscardHalfpairs} reduces to
\begin{equation*}
  R = \frac{k}{n} + \frac{1}{n2^{k-1}} \sum_{i=0}^{\lfloor n/2\rfloor} i A_{2i}.
\end{equation*}
We note that in both cases, dark counts within a codeword are guaranteed to be corrected if $d/2>\lceil n/N\rceil$, and superdense coding requires $c\geq n$, where $c$ is the maximum number of time slots that any EPR pair is guaranteed to remain coherent.

Finally, we propose a third approach where the partially transmitted EPR pair is discarded only if the following round contains no qubit transmissions (i.e. if the codeword is the zero codeword). In all other cases, the transmitter will use the partially transmitted EPR pair for superdense coding.
This utilizes the available entanglement, while still keeping the required decoherence times under control.
As we develop the protocol further in Sec.~\ref{sec:corr-single-photon}, we will in fact need every round to contain at least one qubit transmission.
The necessary modifications are explained in the following section, and we postpone the analysis of $R$ and $E$ to that section.

\subsection{Avoiding silent rounds}\label{sec:avod-silent-rounds}
In anticipation of the requirements of our full protocol capable of handling both dark counts and photon losses, we now modify the protocol to disallow completely silent rounds -- i.e. to disallow the zero codeword to be transmitted.
    In creating a codeword, we will utilize \emph{bit-stuffing} (see e.g.~\cite{Tanenbaum11:BitStuffing}) by letting the sender $U_1$ to insert an extra `1' when the next $k-1$ bits of $\mathsf{B}$ are all zeros.
    That is, if the input buffer is $\mathsf{B}=0\cdots 0 b_kb_{k+1}\cdots$, then $U_1$ will produce the codeword corresponding to message $0\ldots 01$ regardless of the value of $b_k$, and the input stream will be $\mathsf{B}=b_kb_{k+1}\cdots$ afterwards.

In terms of the rate, this implies that the qubit transmissions within each round will communicate $k$ bits of information with probability $(2^{k-1}-1)/2^{k-1}$ and $k-1$ bits with probability $1/2^{k-1}$.
For the information transmitted via superdense coding, note that the only change is related to the all-zero message, which would previously have been mapped to the zero codeword. With bit-stuffing in place, this will instead be mapped to the codeword corresponding to the message $0\ldots 01$, which has some weight $w^\ast$.
Recalling that \emph{every} qubit will now be used for superdense coding as described at the end of Sec.~\ref{sec:corr-single-dark}, the rate is now
\begin{align}\label{eq:rateComplicated}
  R &= \frac{1}{n}\left(
      \frac{2^{k-1}-1}{2^{k-1}}k + \frac{1}{2^{k-1}}(k-1)
      + \frac{1}{2^k}\sum_{i=1}^n iA_i
      + \frac{1}{2^k}w^\ast
      \right) \notag\\
    &=\frac{k}{n} + \frac{1}{n2^k}\left(\sum_{i=1}^n i A_i -2\right) + \frac{1}{n2^k}w^\ast \notag\\
    & > \frac{k}{n} + \frac{1}{n2^k}\left(\sum_{i=1}^n i A_i -2\right).
\end{align}

To compute the efficiency $E$ as defined in~\eqref{eq:defEfficiency}, note that
the expected number of bits transmitted per timeslot is
    \begin{equation*}
      k -\frac{1}{2^{k-1}} + \frac{1}{2^k}\!\left(\sum_{i=1}^n iA_i + w^\ast\!\!\right)
      \!> k -\frac{1}{2^{k-1}} + \frac{1}{2^k}\!\left(\sum_{i=1}^n iA_i + 1\!\right)\!\!,
    \end{equation*}
    while the expected number of physical qubit transmissions is
    \begin{equation*}
      \frac{1}{2^k}\left(\sum_{i=1}^n iA_i w^\ast\right) < \frac{1}{2^k}\left(\sum_{i=1}^n iA_i+n\right).
    \end{equation*}
    Combining these, we obtain
    \begin{align}
      E &>\frac{k -\frac{1}{2^{k-1}} + \frac{1}{2^k}\left(\sum_{i=1}^n iA_i + 1\right)}{\frac{1}{2^k}\left(\sum_{i=1}^n iA_i+n\right)}\notag\\[.5em]
      &= 1 + \frac{2^kk-(n+1)}{\sum_{i=1}^n iA_i+n}.
    \end{align}

With these modifications, we still require $d/2>\lceil n/N\rceil$, but a longer decoherence time is necessary. More precisely, the worst case is that $U_1$ transmits a codeword $(1,\ldots,1,0\ldots,0)$ of odd weight. Since the final, partially transmitted EPR pair needs to stay coherent until the end of the next round, we need coherence time at least $c\geq \max\{n,2n-d+1\}=2n-d+1$.

\textit{Example:} To demonstrate how this works, we use the example given in Fig.~\ref{fig:2WayProtocol}. Suppose that $\C$ is the $[7,4,3]$ binary Hamming code. Note that the requirement $d/2=3/2 > \lceil 7/N\rceil$ implies $N\geq 7$. In other words, we cannot allow more than one dark count per codeword. In fact, for the Hamming code of length $n$, this constraint always implies $N\geq n$, so at most one error per codeword is the limit for all Hamming codes. The users $U_1$ and $U_2$ may now protect the transmission from a dark count as follows. Being the transmitting user in this case, $U_1$ starts by encoding blocks of 4 bits into strings of 7 bits via $\C$. Suppose that $U_1$ has the information string
\begin{equation*}
    \mathsf{B} = 10011011100000\cdots.
\end{equation*}
$U_1$ then applies the encoding procedure proposed in Fig.~\ref{fig:blockdiagram} to obtain the following codewords:
\begin{align*}
    m_1 = 1001 &\overset{\C}{\longmapsto} 0011001 = c_1,\\
    m_2 = 1000 &\overset{\C}{\longmapsto} 1110000 = c_2.
\end{align*}
Additionally, $U_1$ reserves the following pairs for superdense coding
\begin{equation*}
    sd_1 = 10, \quad sd_2 = 11, \quad sd_3 = 01.
\end{equation*}
To initiate the protocol, $U_1$ generates an EPR pair. For the first two time slots, since the first bits in $c_1$ are `0's, $U_1$ remains silent. Then, for the following two time slots, $U_1$ will transmit the modified EPR pair containing $sd_1$. Note that only a single `1' remains in the current codeword, meaning that one half of an EPR pair is transmitted by the end of the first codeword. However, since the next codeword is immediately following, and both users have agreed upon the code applied in the protocol, this is not a concern for the acquisition of the encoded bits. Although the information is split into discrete codewords, the communication should be seen as a constant stream of bits, so $U_2$ must simply wait for an appropriate number of qubits to arrive in order to begin decoding.

After 7 time slots, $U_2$ will have received $c_1'$ in its entirety, which might contain a dark count. Assume for example that $c_1'=1011001$. Then, via $\C$, $U_2$ can easily detect and correct the dark count in the first time slot. 

\begin{figure}[t!]
    \centering
    \resizebox{\columnwidth}{!}{
    \begin{tikzpicture}
\pgfplotsset{
        y axis style/.style={
            yticklabel style=#1,
            ylabel style=#1,
            y axis line style=#1,
            ytick style=#1
       }
   }
\begin{axis}[%
name = figLeft,
width=6.0in,
height=2.33in,
at={(0in,0in)},
scale only axis,
xmin=7,
xmax=255,
xtick distance = 25,
minor x tick num = 1,
xlabel style={font=\color{white!15!black}, font=\Large},
xlabel={$n$},
ymin=1,
ymax=1.5,
minor y tick num = 1,
ylabel style={font=\color{white!15!black}, font=\Large},
tick label style={black, semithick, font=\Large},
ylabel={$R$ [info. bit/slot]},
ytick distance = 0.1,
axis background/.style={fill=white},
xmajorgrids,
ymajorgrids,
legend style={at={(0.97,0.03)}, anchor=south east, legend cell align=left, align=left, draw=white!15!black, font=\huge}
]

\addplot [
    domain=7:255, 
    samples=100, 
    color=black, line width=2.5pt,
    mark=*,
]
table {
7   1.0669
15  1.2333
31  1.3387
63  1.4047
127 1.4448
255 1.4686
};

\end{axis}

\begin{axis}[
width=6.0in,
height=2.33in,
at={(0in,0in)},
scale only axis,
axis y line*=right,
axis x line=none,
xmin=7,xmax=255,
ymin=1, ymax=255,
ylabel=$N$,
y axis style=brightRed,
tick label style = {font=\Large},
ylabel style = {font=\Large},
]

\addplot[
domain=7:255,
samples=100,
color=brightRed,
line width=2.5pt,
mark=*,
]
table {
7   7
15  15
31  31
63  63
127 127
255 255
};

\end{axis}

\end{tikzpicture}





    }
    \caption{Rate (black) and lower bound on $N$ (red) for the protocol capable of correcting dark counts using the $[2^m-1,2^m-m-1,3]$ Hamming code, under the assumption that no more than a single dark count occurs for every $N$ time slots.}
    \label{fig:DC_Hamming}
\end{figure}
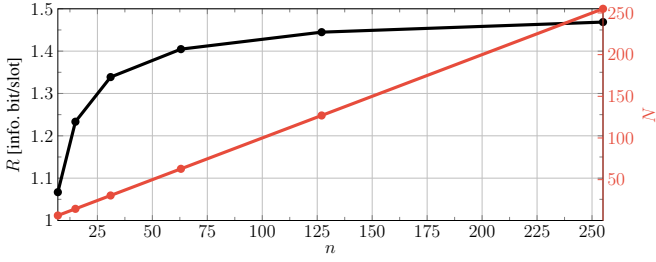

As the weight distribution of binary Hamming codes can be recursively generated, they offer a good baseline for treating the performance of this error-correcting variant. The rate provided by the first few Hamming codes can be seen in Fig.~\ref{fig:DC_Hamming}.

\subsection{Correcting dark counts and photon losses simultaneously}\label{sec:corr-single-photon}
We now consider the case where photon losses may happen.
 
These are more severe than dark counts, as they affect both the sequence of received qubits and the information embedded by superdense coding. Namely, if one half of an EPR pair is lost, the receiver cannot measure it to recover the bits from superdense coding, so these bits are lost unless some redundancy is added.
We attempt to mitigate this threat by adding another layer of error correction to this embedded information, yet again assuming the potential occurrence of a single photon loss during $N$ time slots.
We note, however, that both photon losses and dark counts correspond to (classical) bit errors in the sequence of qubit transmissions. Hence, we can actually relax our assumption such that any $N$ time slots contain at most one photon loss \emph{or} one dark count.

As in Section~\ref{sec:corr-single-dark}, we propose to encode the qubit sequence using a code $\Cout$ with parameters $[\nout,\kout,\dout]$. Similar arguments then ensure that information from $\mathsf{B}$ can be recovered in the presence of errors.

\begin{figure}
    \centering
    \includegraphics[width=.95\linewidth]{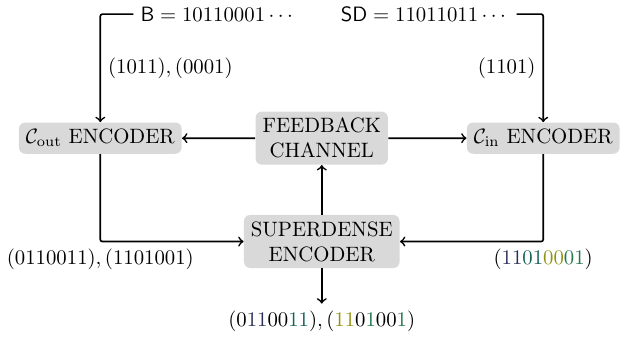}
    \caption{Illustrative example of the idea behind the encoding scheme utilizing two error-correcting codes. Information is divided into two separate strings, $\mathsf{B}$ and $\mathsf{SD}$; the former is fed into the encoder using $\Cout$, and the latter is fed into the encoder using $\Cin$. The codeword produced by $\Cin$ is then used as bits to encode via superdense coding as in Fig.~\ref{fig:blockdiagram}.}
    \label{fig:blockDiagram2}
\end{figure}

To protect the information from $\mathsf{SD}$, user $U_1$ will encode it using an $[\nin,\kin,\din]$ code $\Cin$ before applying superdense coding.
In this way, the transmissions of qubits form codewords of $\Cout$, and each pair of qubits hold two bits from a codeword of $\Cin$.\footnote{While we call these `inner' and `outer' codes, they are not directly related to concatenated codes in classical coding theory.}
A photon loss corresponds to an error in $\Cout$ as well as two erasures in $\Cin$.
Note, however, that `erasure' in this context only aligns with the classical interpretation (i.e. a symbol that is certainly erroneous) if the codeword from $\Cout$ is decoded successfully. Otherwise, the receiver will conclude that $\Cin$ contains erasures but at the wrong positions.
The scheme and the interaction between $\Cout$ and $\Cin$ is illustrated in Fig.~\ref{fig:blockDiagram2}.

The composition of an inner and an outer code may invite problems if the two codes are not matched to each other.
In particular, one may need several successive transmission rounds to obtain $\nin$ qubit transmissions. The exact number depends on the weight distribution of codewords in $\Cout$.
In this way, it is necessary to choose parameters such that (i) $\Cout$ can correct errors occurring within a single transmission round, and (ii) $\Cin$ can correct erasures accumulating across multiple successive transmission rounds.

The first point can be expressed as
\begin{equation}\label{eq:requirementDout}
  \dout/2>\lceil \nout/N\rceil
\end{equation}
as in Section~\ref{sec:corr-single-dark}. For the second point, note that a single codeword of $\Cin$ can span over no more than  
$\lceil (\nin-1)/\dout\rceil+1$ rounds 
(recalling that our bit stuffing strategy rules out completely silent rounds), corresponding to $\nout(\lceil (\nin-1)/\dout\rceil +1)$ time slots.
To see why this is the case, note that the worst case is an inner codeword where either the first or last entry belongs to its own, separate outer codeword. In that case, the remaining $\nin-1$ bits must be spread across the neighbouring outer codewords, which are in the worst case of weight $\dout$.
By dividing this with $N$ and rounding up, we obtain a bound on the number of photon losses that may affect this codeword. Each photon loss corresponds to two erasures, so we obtain the requirement
\begin{equation}\label{eq:requirementDin}
  \din > 2\left\lceil \frac{\nout(1+\lceil (\nin-1) / \dout\rceil)}{N}\right\rceil.
\end{equation}
In other words, the proposed scheme will guarantee correction of a single error if code parameters are chosen such that~\eqref{eq:requirementDout} and \eqref{eq:requirementDin} are satisfied.

The transmission rates of this scheme can be obtained fairly easily by adjusting those of Section~\ref{sec:corr-single-dark} appropriately.
For instance, the rate in~\eqref{eq:rateComplicated} is modified to
\begin{equation}\label{eq:rate_pl}
  R>\frac{1}{\nout}\left(
    \kout - \frac{1}{2^{\kout-1}} + \frac{\kin}{\nin 2^\kout}\sum_{i=1}^\nout iA_i
  \right),
\end{equation}
when treating photon losses as well.
Similarly, the efficiency $E$ of the full protocol is
\begin{align*}
  E &>\frac{\kout-\frac{1}{2^{\kout-1}} + \frac{\kin}{\nin 2^\kout}\left(
        \sum_{i=1}^\nout iA_i +1
      \right)}{\frac{1}{2^\kout}\left(\sum_{i=1}^\nout iA_i +\nout\right)}\\[.5em]
    &=1 + \frac{2^\kout \kout - \frac{\kin}{\nin}(\nout-1)-2}{\sum_{i=1}^\nout iA_i +\nout}.
\end{align*}

\textit{Example:} Let $\Cout$ be the binary Hamming code with parameters $[7,4,3]$, and let $\Cin$ be the $[8,4,4]$ extension of $\Cout$ constructed by adding a parity bit. We note that $\Cin$ is self-dual and has the parity check matrix
\begin{equation}
    H = \begin{bmatrix}
    1 & 0 & 0 & 0 & 1 & 1 & 0 & 1\\
    0 & 1 & 0 & 0 & 0 & 1 & 1 & 1\\
    0 & 0 & 1 & 0 & 1 & 1 & 1 & 0\\
    0 & 0 & 0 & 1 & 1 & 0 & 1 & 1
    \end{bmatrix}.
\end{equation}
Note also that Eqs.~\eqref{eq:requirementDout}-\eqref{eq:requirementDin} imply that $N\geq 21$ should be satisfied, meaning that we assume no more than one error for every three codewords transmitted. Suppose $U_1$ has the following strings they wish to transmit
\begin{align*}
    \mathsf{SD} &= 11011011\cdots \\
    \mathsf{B}  &= 10110001\cdots 
\end{align*}
First, $U_1$ will encode $4$ bits from $\mathsf{SD}$ using $\Cin$, which gives the codeword
\begin{equation*}
    sd_1 = 1101 \overset{\Cin}{\longmapsto} 11010001.
\end{equation*}
Then, $U_1$ will encode $4$ bits  from $\mathsf{B}$ to form the codeword
\begin{equation*}
    b_1 = 1011 \overset{\Cout}{\longmapsto} 0110011
\end{equation*}
to be sent in the first round. Since $b_1$ has weight $4$ corresponding to $2$ EPR pairs, $U_1$ will use superdense coding to encode the first half of $sd_1$ onto these two pairs.

During this process, $U_2$ will receive qubits from $U_1$, but suppose that a photon loss causes a qubit to be lost in transmission. That is, assume for instance that $U_2$ receives $\hat{b}_1=0100011$ instead of $b_1$. Then by the error-correcting properties of $\Cout$, $U_2$ can recover the lost (classical) bit to restore $b_1$. Additionally, this reveals that the lost qubit belonged to the first EPR pair. Thus, $U_2$ knows that $sd_1$ has erasures on the first two indices, but cannot correct these until the remaining bits of $sd_1$ arrive as part of another codeword.

In that regard, $U_1$, as before, is transmitting $4$ bits from $\mathsf{B}$ encoded as:
\begin{equation*}
    b_2 = 0001 \overset{\Cout}{\longmapsto} 1101001.
\end{equation*}
This time, no further encoding is done of $\mathsf{SD}_1$ since there are still $4$ bits remaining from the codeword $sd_1$, and this matches the $4$ `1's in $b_2$. As before, $U_1$ encodes the information in the qubit transmissions and superdense coding.
Now, when $U_2$ receives this information, it can form $b_2$ and the full codeword $sd_1$ apart from the erasures from the previous transmission. That is, $U_2$ holds $\widehat{sd}_1=\mathsf{X}_1\mathsf{X}_2010001$, where $\mathsf{X}_1$ and $\mathsf{X}_2$ are erasures. Now, since $Hx^T=0 \Leftrightarrow x\in\Cin$, we can solve the linear system
\begin{equation*}
  \begin{bmatrix}
    1 & 0 & 0 & 0 & 1 & 1 & 0 & 1\\
    0 & 1 & 0 & 0 & 0 & 1 & 1 & 1\\
    0 & 0 & 1 & 0 & 1 & 1 & 1 & 0\\
    0 & 0 & 0 & 1 & 1 & 0 & 1 & 1
  \end{bmatrix} \begin{bmatrix}
    \mathsf{X}_1\\
    \mathsf{X}_2\\
    0\\
    1\\
    0\\
    0\\
    0\\
    1
  \end{bmatrix} = 0
\end{equation*}
to obtain the unique solution $\mathsf{X}_1 = \mathsf{X}_2 = 1$.
Thus, $U_2$ has successfully restored $sd_1$.

In the example, encoding is made slightly easier due to the input stream being split into $\mathsf{B}$ and $\mathsf{SD}$. In practice, this would have to be done continuously as indicated in Fig.~\ref{fig:blockdiagram}, where the output of the outer encoder is used to determine how many bits to allocate for superdense coding.

\subsection{Optimizing transmission structure} When implementing several layers of error correction to protect against photon losses, we want to choose code parameters in a way that optimizes the structure of transmission rounds. For instance, an unoptimized structure might lead to several rounds of transmission being required before a full codeword from the inner code is obtained.
In turn, this increases the number of potential errors.
One proposal for optimizing the structure is to let the
weight distribution of the outer code align with the block length of the inner code in such a way that codewords are always guaranteed to be completely transmitted within a single round of communications.

In particular, the class of Hadamard codes has parameters $[2^k, k, 2^{k-1}]$ with all non-zero codewords having Hamming weight $2^{k-1}$. Thus, an inner code of length $2^{k-1}$ would ensure that the number of EPR pairs transmitted in a codeword of the outer code matches exactly the number of EPR pairs required by a codeword of the inner code.

We want the parameters to be large enough for the rate to be of any practical value, while keeping them sufficiently small for the qubits to remain coherent throughout the communication protocol. Regarding performance metrics, the rate of the protocol, utilizing a Hadamard code of length $n=2^k$ as outer code and a code of length $2^{k-1}$ as inner code, is given by
\begin{equation*}
    R = \frac{1}{n}\cdot \left(k+\frac{(n-1)R_{\text{in}}}{2} \right),
\end{equation*}
where $R_{\text{in}}=\kin/\nin$ is the rate of the inner code.

Note that for the composition of outer and inner codes to work as intended, the inner code must have block length half of that of the outer code, and it must have minimum distance 3 or greater, since we need to be able to correct 2 erasures. With these constraints, we then want to find the largest message length for the inner code, as this will yield the best possible rate. 
Moving forward, we see that for $k=4$, we obtain the $[16,4,8]$ Hadamard code. Thus, we are searching for an inner code with $n=8$ and $d\geq 3$ and with the greatest message length possible.  
One suggestion is to construct a code by taking the Plotkin sum of the length 4 repetition code and its dual code, that is, $\C=R_4^\perp \oplus R_4$, where $R_n$ denotes the $[n,1,n]$ repetition code.  
The construction of these inner codes can be done recursively, that is, for $k=3$, the inner code is $R_4$, and for $k>3$ the inner code of length $n=2^{k-1}$ can be computed as
\begin{equation}\label{eq:Hadamard-inner-code}
    \C_n = R_{n/2}^\perp \oplus \C_{n/2}. 
\end{equation}
Although not very favorable performance-wise (see Fig.~\ref{fig:hadamard_rate}), this choice is structurally sound and ensures that the number of EPR pairs transmitted in a round always aligns with a codeword from the inner code. In general, for the $[2^k,k,2^{k-1}]$ Hadamard code, the corresponding inner code will have parameters $[2^{k-1},2^{k-1}-k,4]$.

\section{Modelling error probabilities}\label{sec:modelling-error-prob}
To provide context to the value of the bound $N$ on the frequency of errors, we assess the probability of both dark counts and photon losses, taking hardware performance and distance between nodes as well as the duration of each time slot into account. 

In this model, we assume the users to be equipped to transmit photons in the infrared C-band. Specifically we will assume that photons are generated with a wavelength of 1550~nm, as this corresponds to the absorption minimum of silica-based optical fibre. This in turn provides a very low attenuation loss of $\alpha=0.2$ dB/km \cite{Luo22}. Currently, there is a great push towards even lower attenuation loss by using hollow-core optical fibers (HCFs) to allow the photons to mainly travel through air, with recent works reaching attenuation loss as low as $0.05$ dB/km \cite{Petrovich2025,Zou24,Sakr2020}.

To evaluate the probability of dark counts, we must make reasonable assumptions about the background count rate. That is, the total number of counts per second (cps) that is not intended by the transmitter. This includes intrinsic dark counts caused by the photon detector erroneously heralding a non-existing photon, stray photons from the background, blackbody counts caused by thermal radiation of components, or Raman scattering. This varies greatly across different detection equipment. For superconducting nano-wire single photon detectors (SNSPDs), the dark count rate can range between $1\leq d \leq 100$~counts per second (cps) \cite{Gourgues19}, with some experiments having reached rates as low as $d=1.6\times 10^{-2}$~cps \cite{Taylor23}. For our model, we assume that dark count events follow a Poisson distribution with rate $d$. Then the probability of having a dark count in a time slot is
\begin{equation*}
    p_{\mathrm{dc}} = 1 - \mathrm{e}^{-d\tau},
\end{equation*}
where $\tau$ is the duration in seconds of a time slot. Regarding photon losses, we use the photon survival probability based on the attenuation loss as well as the total length, $L$, of the optical fibre. Thus, the probability that a single photon survives the transmission is
\begin{equation*}
    p_{\mathrm{ps}} = 10^{-\alpha L/10}.
\end{equation*}
We also consider the efficiency of the detector, which provides the probability that the detector successfully heralds the detection of a transmitted photon, excluding dark counts, denoted $E_{\det}$. The probability that a photon survives the transmission and is successfully detected is then simply $\eta=p_{\mathrm{ps}}\cdot E_{\det}$. A photon loss occurs exactly when this fails, 
which happens with the probability
\begin{equation*}
    p_{\mathrm{pl}} = 1-\eta.
\end{equation*}

Recall that the length of a single frame of communication is $\nout$ time slots. Consider the random variable $\mathsf{T}$ representing the number of errors throughout the frame. Consider a codeword $c\in\Cout$ with $w(c)=i$ for some $i$. Then, assuming that errors are independent, the probability of having exactly $r$ errors given the codeword $c$ can be calculated combinatorially as in Eq.~\eqref{eq:prob1}.
\begin{table*}
\centering
\begin{equation}\label{eq:prob1}
    \Pr\left(\mathsf{T}=r\,|\, w(c)=i\right) = \sum_{j=\max(0,r-i)}^{\min(r,\nout-i)} \binom{n-i}{j}p_{\text{dc}}^j(1-p_{\text{dc}})^{n-i-j} \binom{i}{r-j}p_{\text{pl}}^{r-j}(1-p_{\text{pl}})^{i-r+j}.
\end{equation}
\end{table*}
Averaging over the entire code, and using $|\Cout|=2^{\kout}$, we obtain 
\begin{equation*}
    \Pr(\mathsf{T}=r) = \frac{1}{2^{\kout}}\sum_{i=0}^{\nout} A_i\Pr(T=r\,|\,w(c)=i).
\end{equation*}

Suppose that the protocol can mitigate at most $t$ errors. Thus, to compute the probability of protocol failure, we simply compute
\begin{equation*}
    P_{\text{fail}} = \Pr(\mathsf{T}>t) = 1 - \Pr(\mathsf{T}\leq t) = 1 - \sum_{r=0}^t \Pr(\mathsf{T}=r).
\end{equation*}

\section{Numerical results}\label{sec:results}
\subsection{Code pair performance}
\begin{figure}[t]
    \centering
    \includegraphics[width=\columnwidth]{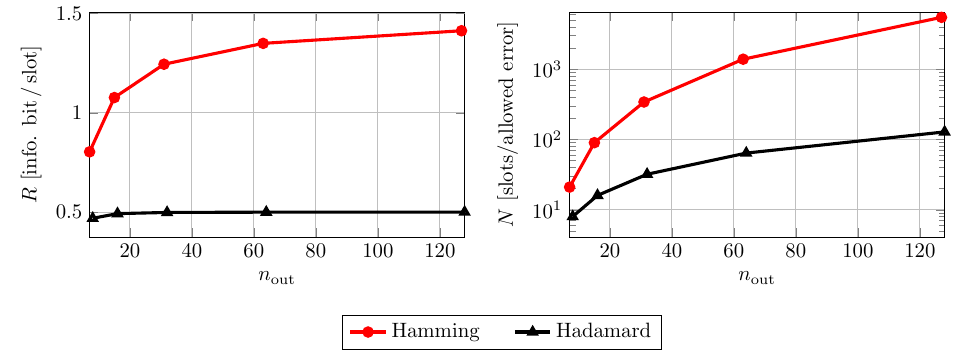}
    \caption{Rate and lower bound on $N$ for the protocol capable of correcting both dark counts and photon losses. Here, ``Hamming'' refers to the scheme using the $[2^m-1,2^m-m-1,3]$ Hamming code as $\Cout$ and the $[2^m,2^m-m-1,4]$ extended Hamming code as $\Cin$, while ``Hadamard'' refers to the scheme using the $[2^k,k,2^{k-1}]$ Hadamard code as $\Cout$ and the code defined in Eq.~\eqref{eq:Hadamard-inner-code} as $\Cin$. The bounds on $N$ are based on Eqs.~\eqref{eq:requirementDout}-\eqref{eq:requirementDin}.}
    \label{fig:hadamard_rate}
\end{figure}

\begin{figure}[t]
    \centering
    \resizebox{\columnwidth}{!}{
    \begin{tikzpicture}
    \begin{axis}[
        width=6.0in,
        height=2.7in,
        at={(0in,0in)},
        scale only axis,
        xmin=0,
        xmax=100,
        minor x tick num = 3,
        xtick distance = 10,
        ymin=0.3,
        ymax=1.2,
        minor y tick num = 1,
        xlabel style={font=\color{white!15!black}, font=\Large},
        ylabel style={font=\color{white!15!black}, font=\Large},
        tick label style={black, semithick, font=\Large},
        xlabel = {Error sparsity parameter $N$},
        ylabel = {Data bit rate $R$},
        legend pos=outer north east,    
        mark=o,
        mark options={
            scale=3,
        },
    ]
        
            \pgfplotstablegetrowsof{code_pairs.dat}
            \pgfmathtruncatemacro{\N}{\pgfplotsretval-1}

        \addplot [
        only marks,
        nodes near coords={
        \pgfmathparse{int(\coordindex+1)}
        \pgfmathresult
        }, 
        every node near coord/.style={
        font=\scriptsize\sffamily\bfseries, 
        anchor=center 
        }
        ] table [x index=1,y index=0] {code_pairs.dat}
        ;

    \end{axis}
\end{tikzpicture}
    }
    \caption{Comparison of performance between a selection of code pairs (see Table~\ref{tab:code_pairs}). We denote by $N$ the lower bound we can achieve on the number of time slots within which a single error may occur, and by $R$ the data rate obtained from Eq.~\eqref{eq:rate_pl}. A legend of the code pairs can be found in Table \ref{tab:code_pairs}.}
    \label{fig:code_pairs}
\end{figure}
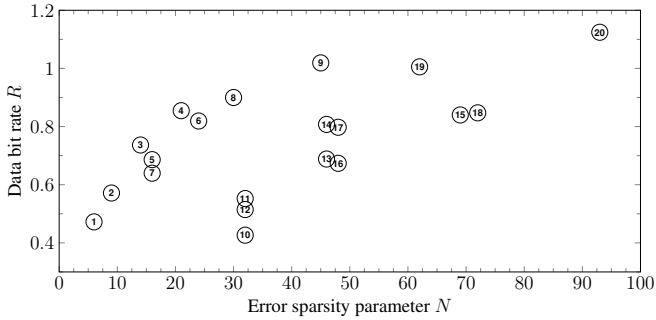

We begin by making a comparison of selected code pairs to visualize the relation between the data rate provided by each pair and the frequency with which we can allow errors, see Fig.~\ref{fig:code_pairs}. From this we can assess that
using Hamming codes provides a much better rate than other codes of similar length that we tested, converging to $1.5$ for increasing sizes of Hamming codes. However, the minimum allowed distance between errors must grow linearly in code length $n$ or equivalently exponentially in $k$. Specifically, for the Hamming code of length $n$, we can allow no more than one error for every $\lceil (n+1)/3\rceil$ codewords. In contrast, when using Hadamard codes the constraints on $N$ from Eqs.~\eqref{eq:requirementDout}-\eqref{eq:requirementDin} imply the lower bound $N\geq 2\nout$, see Fig.~\ref{fig:hadamard_rate}. Thus, we can allow one error in every two codewords, no matter the length. However, due to the way $\Cin$ is constructed, we may tighten this bound to $N\geq \nout$, thus allowing one error in every codeword. We see then that the error-correcting capabilities are much better in the latter case at the cost of data rate.

\begin{figure}[t]
    \centering
    \includegraphics[width=\linewidth]{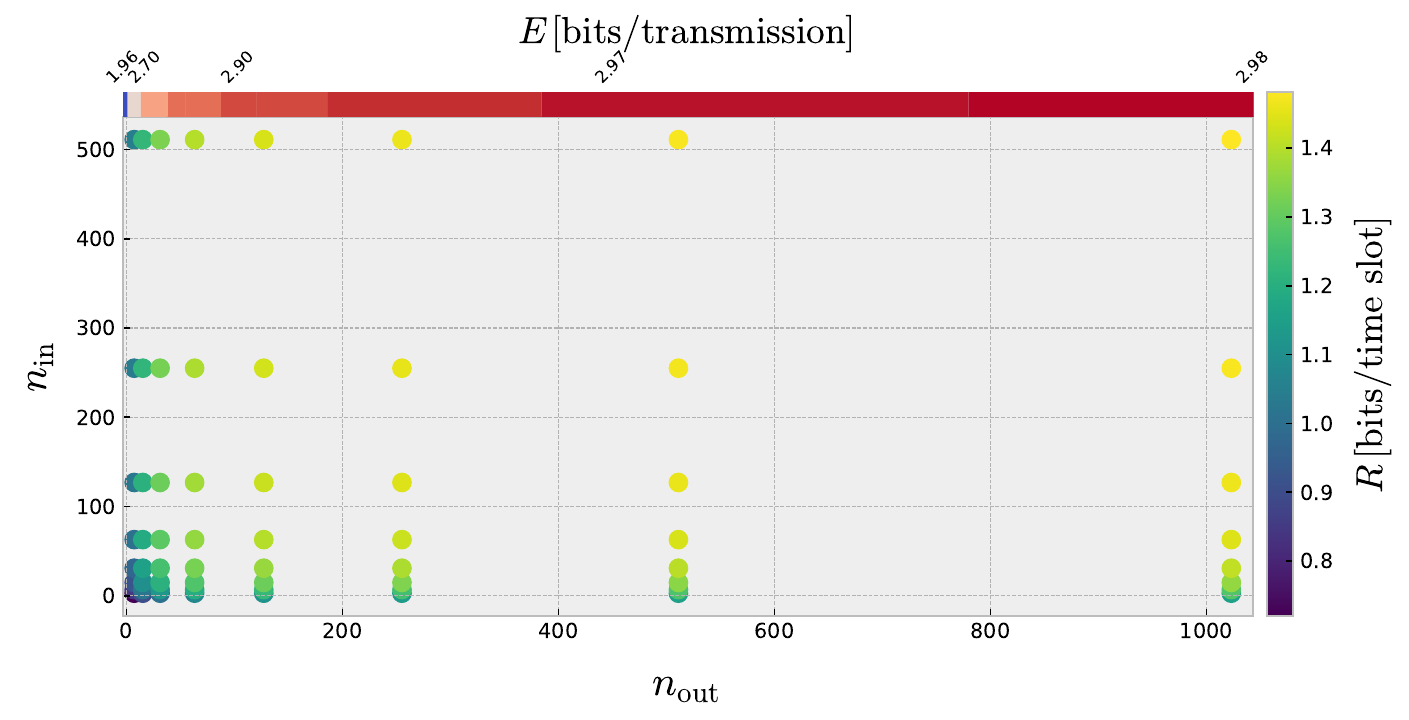}
    \caption{Data rate $R$ and energy efficiency $E$ for all code pairs $(H_{\nout},H_{\nin})$ with $3\leq\nout\leq 1023$ and $3\leq\nin\leq 511$. As $E$ does not depend on $\nin$, it is plotted as a bar parallel to the first axis.}
    \label{fig:HammingRE}
\end{figure}

\begin{figure}[!ht]
    \centering
    \includegraphics[width=0.8\columnwidth]{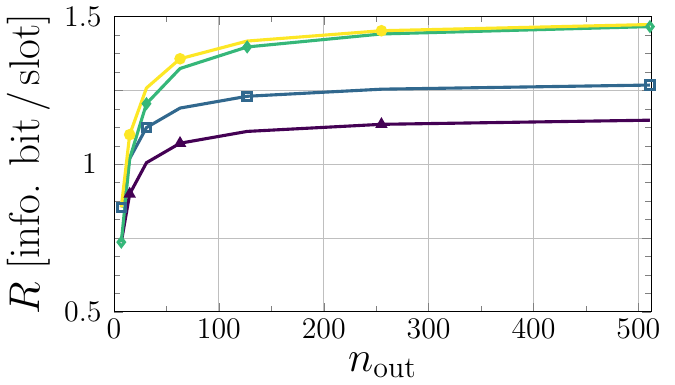}
    \includegraphics[width=0.8\columnwidth]{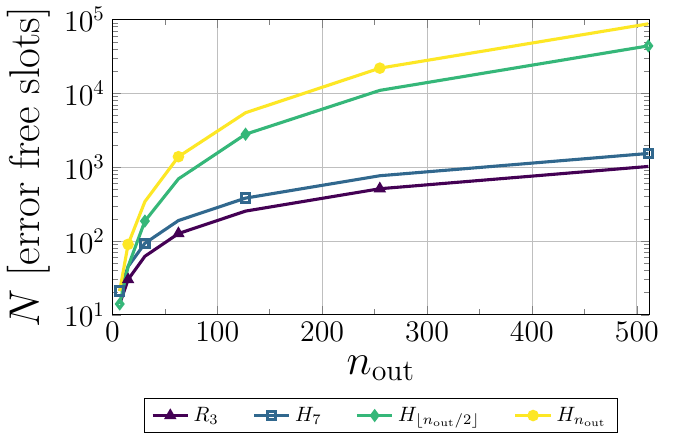}
    \caption{Performance metrics $R$ and $N$ for code pairs $(H_n,\Cin)$ with different inner codes.}
    \label{fig:performanceInner}
\end{figure}

Based on the results from using pairs of Hamming codes, we further explore code pairs of this structure. In Fig.~\ref{fig:HammingRE} we plot the performance of all combinations of Hamming code pairs $(H_{\nout},H_{\nin})$ with $3\leq\nout\leq 1023$ and $3\leq\nin\leq 511$. Although the energy efficiency does not depend on the length of the inner code, both codes contribute to the data rate based on their individual code lengths. To closer inspect the impact of different lengths for the inner Hamming code, we plot both data rate and the minimum number of required error free time slots, $N$, against the outer code length for different inner codes, see Fig.~\ref{fig:performanceInner}. Note that as the length of the inner code approaches that of the outer code, the asymptotic rate approaches the one derived in \cite{JensenEtAl25}. In particular, note that using the Hamming code directly preceding the outer Hamming code in length yields almost the same rate despite having significantly lower demands on the scarcity of errors. The bound on $N$ is, however, still increasing to extreme values.

\begin{table}[!ht]
    \centering
        \caption{Legend of code pairs whose performance metrics $(N,R)$ are plotted in Fig.~\ref{fig:code_pairs}. We denote by $R_n$ the length $n$ repetition code; by $H_n$ the length $n$ Hamming code; by $Had_n$ the length $n$ Hadamard code; and by $G_n$ the length $n$ Golay code. Additionally, $C_1$ is an $[11,7,3]$ code and $C_2$ is a $[12,8,3]$ code \cite{Gra07:Codes}.}
\def\arraystretch{0.5}
\begin{tabular}{cl}
  \toprule
  \textbf{Index} & $\bm{(\Cout,\Cin)}$ \\
  \midrule
  1              & $(R_3,R_3)$        \\
  2              & $(R_3,H_7)$        \\
  3              & $(H_7,R_3)$        \\
  4              & $(H_7,H_7)$        \\
  5              & $(H_8,R_3)$        \\
  6              & $(H_8,H_7)$        \\
  7              & $(H_8,R_4)$        \\
  8              & $(H_{15},R_3)$     \\
  9              & $(H_{15},H_7)$     \\
  10              & $(Had_{16},R_3)$   \\
  \bottomrule
\end{tabular}
\hspace{5mm}
\begin{tabular}{cl}
  \toprule
  \textbf{Index} & $\bm{(\Cout,\Cin)}$ \\
  \midrule
  11             & $(Had_{16},H_7)$   \\
  12             & $(Had_{16},H_8)$   \\
  13             & $(G_{23},R_3)$     \\
  14             & $(G_{23},H_7)$     \\
  15             & $(G_{23},C_1)$     \\
  16             & $(G_{24},R_3)$     \\
  17             & $(G_{24},H_7)$     \\
  18             & $(G_{24},C_2)$     \\
  19             & $(H_{31},R_3)$     \\
  20             & $(H_{31},H_7)$     \\
  \bottomrule
\end{tabular}
\label{tab:code_pairs}
\end{table}

\subsection{Comparison against superdense coding}\label{sec:comp-against-superd}
To evaluate the performance of the proposed protocol, we will use superdense coding on its own as the benchmark.
    That is, we compare against the case where one timeslot is used to transmit one half of an EPR-pair, and the next is used for transmitting two bits using superdense coding. Of course, these bits must still come from an error-correcting code to protect against photon losses and dark counts.
    Since the receiver expects exactly one qubit in each timeslot in this setup, we can assume that any photon loss or dark count is detected by the receiver. In this case, the associated EPR-pair is discarded, and the two bits are treated as erasures.
    Similarly to Sec.~\ref{sec:corr-single-dark}, this means that errors are guaranteed to be correctable if the bits are encoded using an $[n,k,d]$-code with $d/2>\lceil n/N\rceil$.
    To analyse the rate, we split into two cases. First, if $n\leq N$, then the bound is always satisfied for $d=3$. Assuming that the utilized code is maximum-distance separable (MDS), i.e., $d=n-k+1$,  it implies the rate $k/n=(n-2)/n$. This is maximized by maximizing $n$, yielding $k/n\leq 1-2/N$.
    Otherwise, for $n>N$ we once again assume an MDS-code and relax the condition to $d/2>n/N$ to give a bound on the best-case performance. Substituting $d=n-k+1$ and rearranging then gives
    \begin{equation*}
      \frac{k}{n}< 1+\frac{1}{n}-\frac{2}{N} < 1-\frac{1}{N}.
    \end{equation*}
    Since this bound is better than the bound for $n\leq N$, we use it going forward.
    Note that $k/n$ corresponds exactly to the transmission rate $R_{\mathrm{SD}}$ and the efficiency $E_{\mathrm{SD}}$ since two codeword symbols are sent every two timeslots.
    We stress that the bounds $R_\mathrm{SD}=E_{\mathrm{SD}}\leq 1-1/N$ are best-case bounds on the performance of superdense coding. If our proposed protocol exceeds this performance, then it will perform better than superdense coding paired with any error-correcting code.

\begin{figure}[ht]
    \centering
    \includegraphics[width=0.95\linewidth]{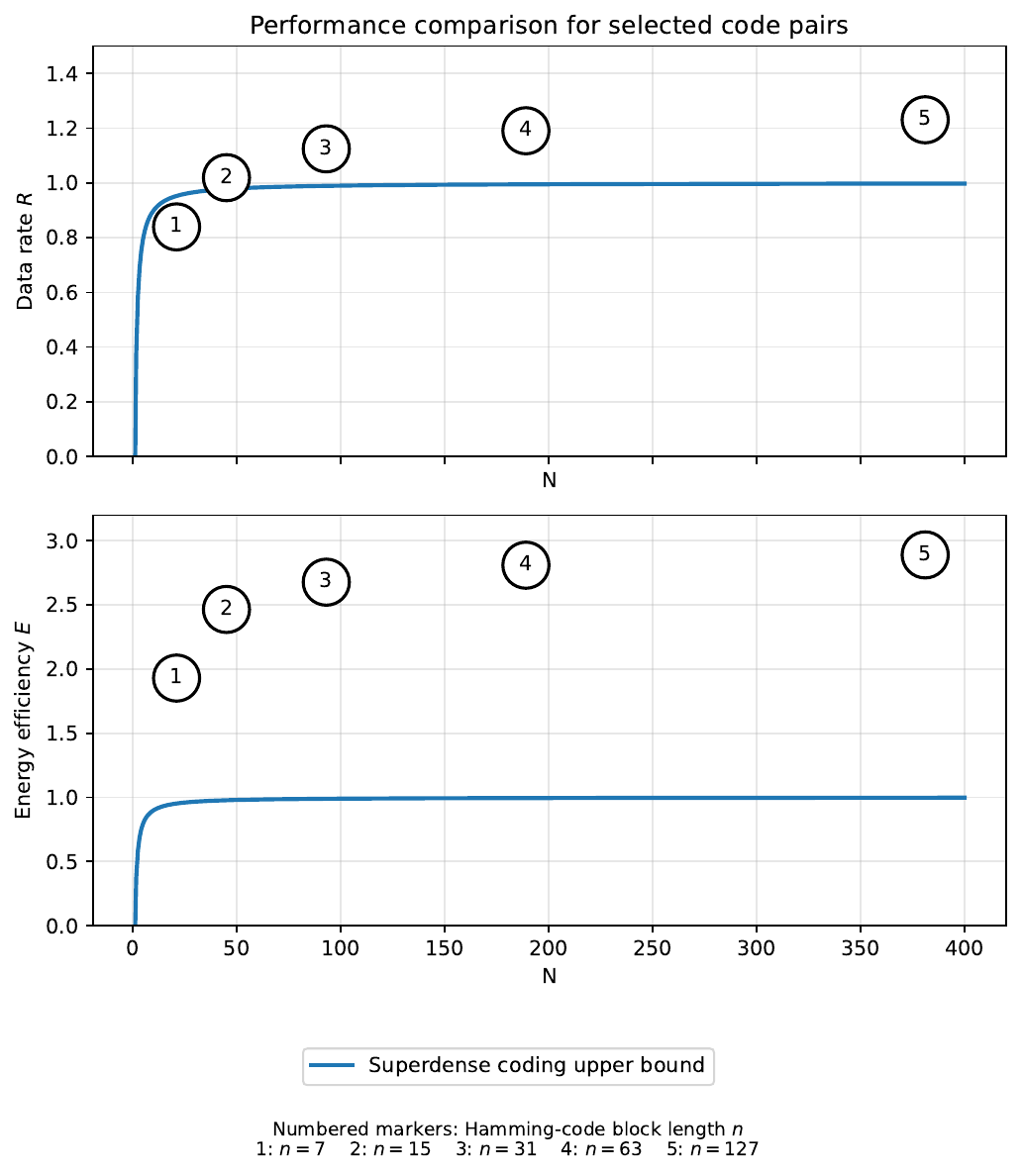}
    \caption{Performance comparison between our proposed protocol and conventional superdense coding protected by classical error correction under the same constraints on the error frequency. The blue curves represent the upper bounds on data rate and energy efficiency, respectively, as derived in Section~\ref{sec:comp-against-superd}. The circles denote selected pairs of Hamming codes employed in our protocol. The first axis describes the bound on $N$ which must be satisfied to reach the corresponding rate and efficiency.}
    \label{fig:performanceCompareSD}
\end{figure}

In Fig.~\ref{fig:performanceCompareSD} we showcase how the protocol performs compared to the superdense coding benchmark. Based on the patterns in performance yielded by utilizing Hamming codes (see e.g. Fig.~\ref{fig:code_pairs}), we base the comparison on code pairs $(\Cout,\Cin)=(\mathrm{Hamming}(n),\mathrm{Hamming(7))}$ with $n=2^m-1$ for $m=3,\ldots,7$. We compare the data rate $R$ and the energy efficiency $E$ to the corresponding metrics $R_{\mathrm{SD}}$ and $E_{\mathrm{SD}}$, respectively, as defined above. From this comparison we see that our protocol performs at a higher data rate from the $[15,11,3]$ Hamming code and onwards. Notably, our protocol performs at a significantly higher energy efficiency that converges to a 200\% increase compared to the superdense coding baseline. We emphasize that the curves based on $R_{\mathrm{SD}}$ and $E_{\mathrm{SD}}$ should be considered the upper bounds on the performance of the superdense coding procedure.

\subsection{Error probabilities}
\begin{figure}
    \centering
    \includegraphics[width=0.95\linewidth]{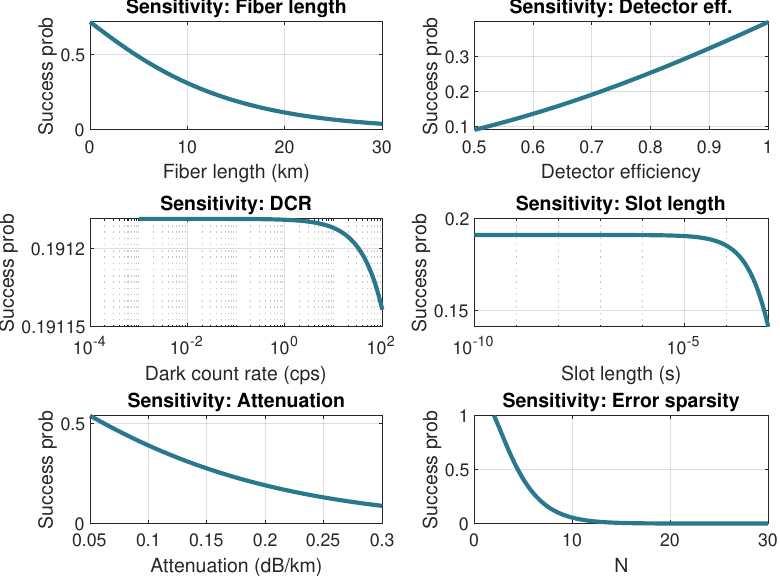}
    \caption{Sensitivity plots of the parameters in the protocol error model.}
    \label{fig:sensitivity}
\end{figure}

Next, we assess the failure rate of the protocol. We begin by simply examining the probability of failure for a single round of communication. For this purpose we assume that $\Cout$ is the $[7,4,3]$ Hamming code. In Fig.~\ref{fig:sensitivity}, we assess the impact of different parameters on the final success rate of rounds of communication. Obviously, the impact of the dark count rate and the slot length is negligible over wide range of the values of these parameters, in comparison to the remaining parameters assessed. By far the greatest challenge is the distance between the users as the attenuation increases exponentially with fiber length. Although the success rate increases with higher detector efficiency, these improvements are dominated by the added failure rate from the increasing fiber length. 

\begin{figure}
    \centering
    \includegraphics[width=0.95\linewidth]{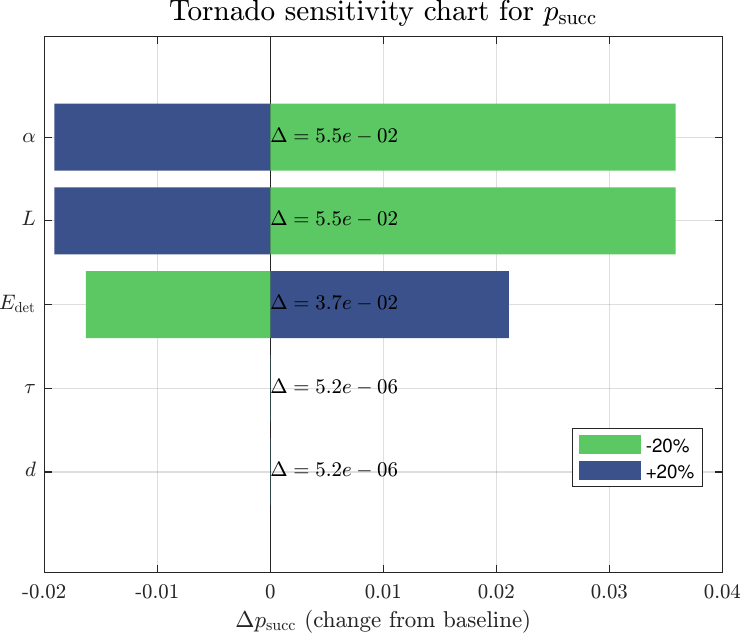}
    \caption{Tornado diagram showing the impact on the protocol success probability from perturbations in parameters values of 20\%. Baseline values for the parameters: $\alpha=0.2$, $L=30$, $E_{\det}=0.7$, $\tau=10^{-6}$, $d=100$. The $\Delta$ values on each bar denote the total contribution to the change in success rate across the entire change of each parameter.}
    \label{fig:tornado}
\end{figure}

\begin{figure}
    \centering
    \includegraphics[width=0.95\linewidth]{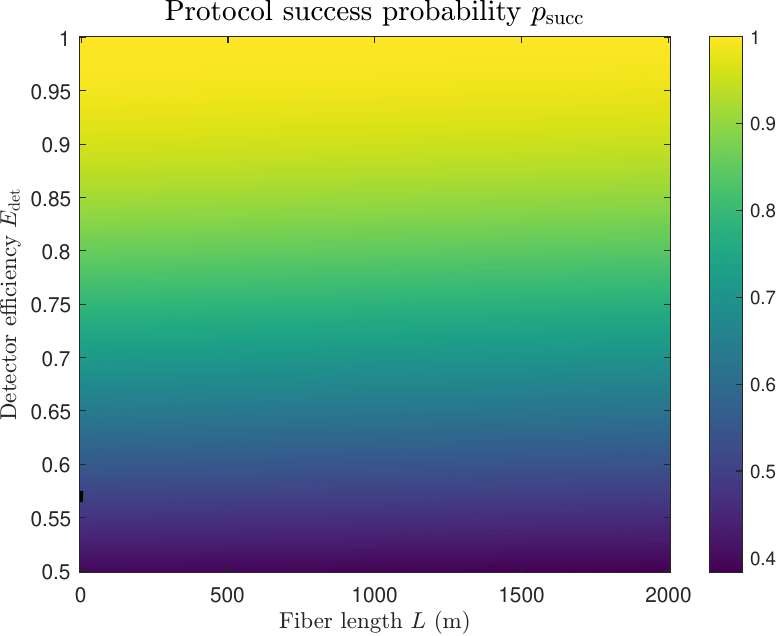}
    \caption{Heat map showing probability of success within a single round of communication across varying fiber length and detector efficiency. For this plot, we have applied $\alpha=0.04$, $d=0.01$, and $\tau=10^{-9}$.}
    \label{fig:heatmap}
\end{figure}

\begin{figure}
    \centering
    \includegraphics[width=0.95\linewidth]{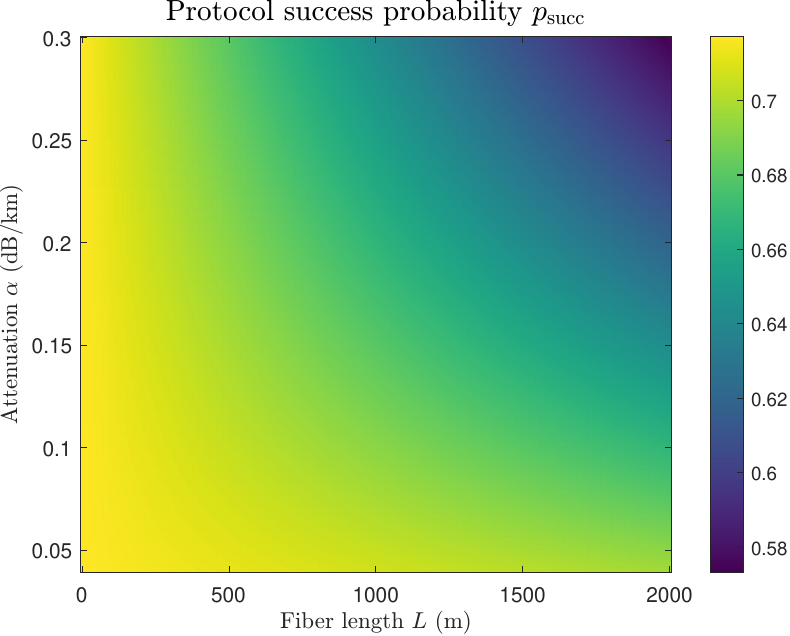}
    \caption{Heat map showing probability of success within a single round of communication across varying fiber length and attenuation. For this plot, we have applied $E_{\det}=0.95$, $d=0.01$, and $\tau=10^{-9}$.}
    \label{fig:heatmap2}
\end{figure}

A comparison of the parameters can also be seen in Fig.~\ref{fig:tornado}. From this, it is very clear that both slot length and dark count rate can be disregarded when compared to the impact in error probability from attenuation and fibre length alone. Since the fiber length is the most sensitive parameter, we produce a heat map between this and the detector efficiency to visualize a region in which the probability of failure is lowest, see Fig.~\ref{fig:heatmap}. From this we observe that for short distances, a sufficiently high detector efficiency allows the protocol to deliver high reliability. Similarly, we produce a heat map evaluating the relation between fiber length and the attenuation loss $\alpha$, with the range $0.04-0.3$ chosen based on the current records from \cite{Zou24} for the lower bound, see Fig.~\ref{fig:heatmap2}.  

With the model parameters used so far, the performance of the proposed approach is somewhat limited. To provide better context, we compute the probability of failure in a best-case scenario defined as follows: Assume $\alpha=0.04$ as reported in \cite{Zou24} and reduce the fiber length down to $L=1$ km. Further assume $d=0.01$, which is within the range reached in \cite{Taylor23}, and let $\tau=10^{-9}$. Finally, let $E_{\det}=0.95$. Suppose we use the code pair $(H_7,H_3)$, resulting in $N=14$. Using this setup, we obtain $p_{\text{fail}}^N \approx 0.059$. However, as soon as we increase the fiber length to $L=10$ km, we get $p_{\text{fail}}^N \approx 0.238$. 

\begin{figure*}
    \centering
    \includegraphics[width=\textwidth]{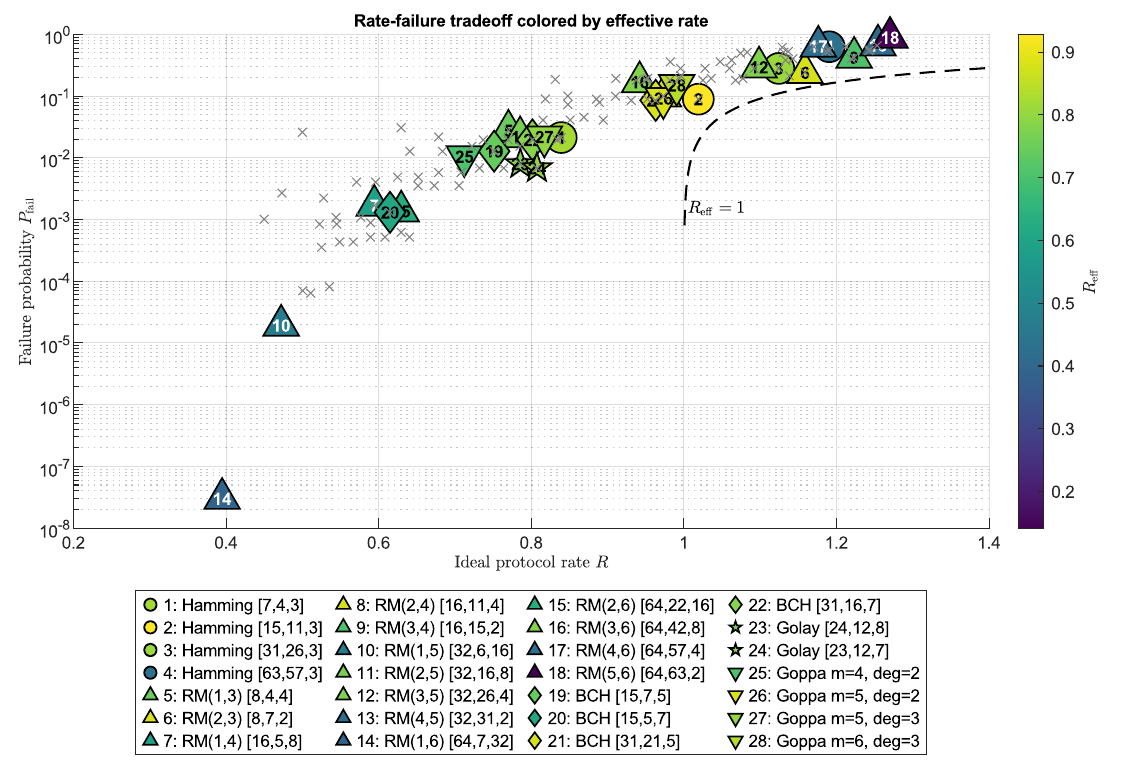}
    \caption{Trade-off between ideal protocol rate $R$ and protocol failure probability $P_{\text{fail}}$ under photon-loss and dark-count noise. Marker colors indicate the resulting effective rate $R_{\text{eff}}=R(1-P_{\text{fail}})$. Numbered markers represent selected structured code families, including Hamming, Reed-Muller, BCH, Golay, and Goppa codes, while gray crosses indicate best-known binary linear codes obtained from the GUAVA/CodeTables \cite{Gra07:Codes} database for $n\in\{7,8,15,16,31,32,63,64,127,128,255,256\}$ and $k\in[\max(2,\lfloor n/4\rfloor),n)$, truncated to 100 codes. The dashed curve denotes the boundary $R_{\text{eff}}=1$. For this comparison, we have fixed $\Cin$ to be the Hamming $[7,4,3]$ code.}
    \label{fig:codeGrid}
\end{figure*}

Clearly, the main contributor to this detriment is the fiber length. Thus, we base the remaining results around a fiber length of at most 2 km. We still use the optimized parameters from before, that is, $\alpha=0.04$, $\tau=10^{-9}$, $d=0.01$, and $E_{\det}=0.95$. The following is an exploratory comparison of performance for several classes of well-known error-correcting codes. The baseline of codes consist of the best-known codes for an array of parameters $n$ and $k$ obtained from \cite{Gra07:Codes}. Additionally, codes with well-known structure such as Hamming, BCH, Golay, Reed-Muller, and Goppa codes, are compared in terms of their effective rate $R_{\text{eff}}=R\cdot(1-P_{\text{fail}})$. This comparison is depicted in Fig.~\ref{fig:codeGrid}.

Obviously, for most of the considered configurations, the failure rate of the proposed protocol is non-negligible. This type of challenge is frequently encountered in, e.g., wireless communications, and necessitates introduction of mechanisms that will allow for detection of failures and their corrections. While the latter can be done via codeword retransmissions, the former could be addressed by appending a sort of a (frame) check sequence that is created on the basis of the transmitted codeword, similarly to cyclic-redundancy checks (CRCs) that are standardly used on the link layer. This could be complemented by adding a synchronization sequence that will ensure delimitation among the codewords, if needed in practice, etc. In other words, the reliability of the solution can be brought to a desired level by developing a full-fledged communication protocol based on the standard communication engineering practice.

\section{Discussion}
\label{sec:discussion}
The presented results indicate that the proposed scheme is primarily suited for short-distance quantum links under the considered physical assumptions. In particular, the protocol exhibits the strongest performance in regimes where attenuation remains sufficiently limited, which in our simulations corresponds to fiber lengths on the order of a few kilometers or less. While this distance regime may appear restrictive in the context of large-scale quantum networks, it remains relevant for several practical scenarios \cite{Caleffi2024,Barral2025}. One example is short-range optical interconnects in distributed computing environments such as data centers, where physically separated computational resources communicate through optical fiber infrastructure over comparatively small distances \cite{Cacciapuoti2026}. In such settings, the reduced propagation distance may help mitigate the dominant impairment caused by attenuation, while the high communication throughput of the proposed protocol could make efficient use of available quantum resources.

Furthermore, short-distance quantum links are expected to play an important role in modular quantum computing architectures and local quantum networking, where nearby quantum processors exchange both classical and quantum information \cite{Awschalom2021,Monroe2014}. The integration of entanglement distribution directly into the communication protocol may therefore provide practical advantages in systems where repeated short-range transmissions occur frequently.

For communication across significantly longer distances, additional infrastructure would likely be required. In particular, satellite-assisted quantum communication and quantum repeater networks may provide mechanisms for extending the operational range of such protocols \cite{deForgesdeParny2023}. However, incorporating these technologies would introduce additional physical considerations beyond the scope of the present work.

\section{Conclusions and future work}\label{sec:conclusion}
In this work, we have proposed a classical communication protocol that combines classical error-correcting codes, entanglement distribution, and superdense coding. By encoding information both in the positions of qubit transmissions and through superdense coding, the proposed scheme integrates the generation and distribution of entanglement directly into the communication process. Classical error correction is incorporated to mitigate the effects of dark counts and photon losses, allowing the protocol to operate under non-ideal channel conditions.

Our analysis demonstrates that this joint classical-quantum encoding approach can provide performance advantages over conventional superdense coding protected by classical error correction. Under identical constraints on the frequency of errors, suitable choices of error-correcting codes exceed the derived upper bounds for conventional superdense coding in terms of both data rate and energy efficiency. In particular, the considered Hamming-code configurations exceed the superdense-coding rate bound from the $[15,11,3]$ code onward, while the energy-efficiency gain approaches a factor of three.

The physical error model nevertheless shows that attenuation places a strong restriction on the operational range of the protocol. Under the considered parameters, the most favorable operating regime is therefore found for fiber links on the order of a few kilometers or less. Rather than targeting long-distance quantum communication, the proposed scheme is consequently particularly applicable to short-range optical links, including quantum data centers, modular quantum-computing architectures, and other local quantum-networking scenarios. 

Several aspects remain to be addressed before the proposed scheme can be considered as a complete communication protocol. In particular, the present work assumes perfect EPR-pair generation and considers a simplified physical error model. Future work should therefore investigate the impact of imperfect entanglement generation and additional sources of quantum-channel noise. Further, practical implementations would require mechanisms for detecting failed transmission rounds, maintaining synchronization, and retransmitting corrupted data. Developing these mechanisms, together with an experimental evaluation of the proposed protocol in a short-range optical setting, represents a natural direction for future work.



\bibliographystyle{IEEEtran}
\bibliography{refs}

\end{document}